# The independence of Central Banks, a *reductio ad impossibile*

Ion Pohoață[a], Delia-Elena Diaconașu[b] and Ioana Negru[c]*

[a]*Department of Economics and International Relations, Faculty of Economics and Business Administration, Alexandru Ioan Cuza University of Iaşi, Iaşi, Romania*
[b]*Department of Social Sciences and Humanities, Institute of Interdisciplinary Research, Alexandru Ioan Cuza University of Iaşi, Iaşi, Romania*
[c]*Department of Management, Marketing, Business Administration, Faculty of Economic Sciences, Lucian Blaga University of Sibiu, Sibiu, Romania*

*corresponding author: delia.diaconasu@uaic.ro

# The independence of Central Banks, a *reductio ad impossibile*

**Abstract:** This paper testifies to the fact that the independence of the Central Banks, as stated by its founding fathers, is nothing more than a chimera. We demonstrate that the hypothesis "inflation is a purely monetary phenomenon" does not support the plea for independence. Moreover, we show that the conservative central banker, the imaginary Principal-Agent contract, the alleged financial autonomy, just like the ban on budgetary financing, are all arguments that lack logic. We equally show that the idea of independence is not convincing because its operational toolbox, as well as the system of rules it relies on, lack well-defined outlines.

## 1. Introduction

We argue that the Euclid's *reductio ad absurdum* suggests the veiled drivers encapsulated in the enormous effort worn out while trying to motivate the independence of the Central Banks (CBI).

First, a brief survey of the area is likely to unveil an obvious disproportion in the power balance. On the one hand, there are the positions of those who set the tone, the note and the method – the bankers. On the other hand, there are the economists in academia, overindulgently welcomed in the sphere of analyses, who put forth interesting ideas but yet to be gelled into actual leads to follow. What needs to be done, the path, the normative part, comes from the bank. That is the citadel of banking science, manifestly autonomous, wrapped in elliptical and oracular expressions. A conclave science, closed to foreign interference. Best (2022) notes that central banks' skillful management of "unknown knowns" – the uncertain and complex economy which they do not fully understand – is just to support a drive towards 'scientization'. The famous Maastricht Treaty (Article 107) and the Statute of the ESCB (Article 7) – which protect them from the instructions and advice of others – fit the bankers like a glove. They can work freely on their science. In the name of science and under its cover, they can refuse unfriendly places and voices.

Second, the approaches on the topic of the independence of the Central Banks walk whoever is interested through all the important areas of the economy: inflation, employment, growth, money, credit, prices, budget, finance, value, interest, exchange rate, and so on. As interesting as this journey may be, it is not as compelling in terms of the reasoning in support of the independence (see, e.g., Forder 2005, for a critical and interesting overview of the reasons that led to the rapidity with which central bank independence was adopted). The idea is induced that the CBI originates in weighty areas, that is, areas with unquestioning roots.

Third, supporters of the idea of central bank independence link their discourse to the "time inconsistency" of Kydland and Prescott (1977), which argues that policymakers should commit to a policy *rule* in advance rather than have discretion that suffers from lack of credibility. The implication of this idea, at the normative level, is that some institutions that govern policymaking are better than others at making effective and credible decisions about monetary policy (Tabellini 2005). So, the attention was turned to the Central Bank, known for its fight against the government's inflationary leanings. Barro and Gordon (1983) explicitly considered central bank reputation and credibility in the fight against inflation (in fact, replacing the rule with the *credibility* of the organization). Rogoff (1985) takes one step further in this direction: as an alternative to the idea of reputation, he suggests entrusting the monetary policy to a *conservative*, non-lax banker, one showing a clear commitment to price stability above that of most economic agents. A *delegation of legitimacy* concludes, through Walsh (1995), the anti-time inconsistency cycle. The operation is carried out through an *enticing type of contract* in the form of the *Principal-Agent*. The principal (the government) empowers the Central Bank with monetary policy on efficiency grounds. As initiated by Walsh, but especially by those who patched up the idea, the contract includes *rules* on the Central Banks' budgetary independence, the ban on direct budgetary funding, the structure and duration of mandates, the governor's salary and granting conditions; the goal: price stability. The monopoly on monetary issuance and the status of lender of last resort are not included in the contract; they are considered implicit.

With such roots, with a stubbornness fed from an allegedly own budget, unrivalled, the position of the Central Bank seems to be canonized. This paper aims to prove that, relying on conceptual and institutional laxity, the independence of a Central Bank, as invoked today, is a chimera. In this regard, the problem is tackled in three directions. The first objective is to unveil the weaknesses of its theoretical construct. Here, is considered the hypothesis in which the idea of independence is grounded: inflation is a purely monetary phenomenon. A second objective, derived from the first, translates into an effort to decode the splits between the main circumstances that serve as arguments in imposing the idea of independence, i.e., the conservative central banker; the imaginary Principal-Agent contract; the lender of last resort; the alleged financial autonomy on other people's money; and the dilemmatic ban on the Central Bank budgetary funding. Systematically, the analysis will show that relying on such facts, the founders' arguments limp behind. Finally, this study shows why the independence of the Central Bank cannot be a realistic idea, since its institutional support, as well as its conceptual toolbox, the rules that should rigorously regulate its behavior, are vaguely shaped.

## 2. Doubtful hypothesis in the plea for independence

Friedman's famous dictum (1968, 39), "Inflation is always and everywhere a monetary phenomenon" was successfully exploited in the seminal arguments of CBI in the 1970s and 80s. Regarding this hypothesis, it should be noted that there are two hypostases of economic theory and practice: one that confirms Friedman, and one that refutes his theory.

Inflation is *indeed* a monetary phenomenon as a form of manifestation. One does not find it and it is not noticeable through the explosion of supply or demand for goods or labor, but through the unbridled explosion of prices. Prices not as expressions of production's long-lasting value, but prices expanded as a result of a monetary laxity originating in the works of Menger (2009) and those who followed in his footsteps in the line of thought of the Austrian School. Without stating it directly, the proponents of the Mengerian idea of a neutral currency draw on the fact that the only function of money is that of means of circulation. It can be sent in any amount on the channels of money circulation; it will not influence the real economy. The standard of value appears to be an artefact. This "dauntless" invention, backed by Mises (2013) and Hayek (1967), paved the way for "relaxation", an endearment of today's damaging laxity.

Inflation is *not* a monetary phenomenon in terms of origins and working mechanism. Prices are forms of expression of the forged values, which incur production costs, and which are dimensioned subjectively on the market under the rule of the demand-supply ratio. Here we should rather look at Ricardo (2001) and Marx (1990); at Marshall (2013) and his well-known synthesis; at Sraffa's (1960) "commodities produced by means of commodities"; at Hayek (1967), with the active relationship between price and production structures; and all the other Austrians (e.g., de Soto 2006) who believe that if the production process is a sequence of a circuit that started a long time ago, prices are the ones that determine costs and not the other way around. In other words, a line of thought sensitive to the importance of costs and the amount of money in circulation in price formation.

Why is such an opinion unsuccessful and inapplicable? Because it suggests that the amount of money is not neutral in the process of value creation and that of expressing it through money. In addition, it rejects the idea that inflation stems solely from the movement of money. Not accepting it means defying reality; and yet, this defiance is good for the idea of independence. If the currency is neutral, the Bank can manage its monetary policy with no interference from the outside. In addition, in such circumstances, it is not engaged in responsibilities regarding the dynamics of the real world. The bank aims for inflation, as a gateway to the real economy, but under its shell, the currency is neutral. The same doubtful philosophy holds for targeting prices to which the amount of money does not participate. In this context, none of the supporters of the idea of independence has provided a convincing answer as to why we are targeting inflation and not prices; or vice versa. The sandy ground on which this analysis is built gives room for distrust. How could one remain assured, when the material under analysis itself is deformed and, most often, marginalized? How many of those enthused by the idea of independence still wonder what money really is? What monetary aggregate are they thinking of, intending to move it, manipulating the interest rate? What are they aiming at, $M_1.M_2, ... M_n$ or $L$?

An independent Central Bank's persistence to target inflation – the conventional wisdom lower inflation rate – central bank independence (see, e.g., Cukierman 2008; Carlstrom and Fuerst 2009; Arnone and Romelli 2013; Bodea and Hicks 2015; Garriga and Rodriguez 2020) –, on such shaky ground, is alluring, illusory. Michel Aglietta also believes that on an ambiguous notion, broken apart from the classical discourse—such as the value of currency— it is impossible to build solidly on the idea of independence. Moreover, essentially, if the currency has no bearing on the real world, "the Central Bank's independence would be an empty issue, devoid of significant content" (Aglietta 1992, 16).

Even confined to its function as a means of payment, money can be thought of in any possible way, anything but neutral. However, expressing prices in an active currency would change things. If one links inflation to all its known and unknown determinants, to what is going on in terms of costs, employment policy and the implementation of investments, imports, and exports, etc., what is aimed at through the monetary policy interest rate? And how accurately? Is inflation our problem alone or is it the government's also? Or is it particularly the government's? If we consider the two twin ideas together – prices are expressed in an active currency and inflation cannot be a purely monetary phenomenon – we can infer two consequences: (a) The dynamics of prices does not belong exclusively to the nominal world; (b) The explosion of prices, inflation, does not belong exclusively to this world either. It equally belongs to the real world, and as such, there must be room for other forces to enter this game. The independence of Central Banks alone does not solve the problem of inflation (Klomp and de Haan 2010; Hayo and Hefeker 2002). The independence of Central Banks is not absolutely necessary or sufficient for monetary stability (to reduce inflation). In this regard, Fullwiler and Allen (2007) call into question the very ability of the Fed to target inflation.

One can turn a blind eye to the fact that M is hard to define, and that L escapes the Central Bank's analysis; that the line between monetary and non-monetary assets is difficult to draw; that the money supply can also increase at the expense of public spending policy, the wages; that inflation can be imported, etc., and then declare and legislate the independence of entral Banks. On what? On a monetary policy that operates with a neutral currency and targets a purely monetary phenomenon. Not even Milton Friedman would fit such a logic. And that is because the much-cited idea was taken out of context. Taken out to prove something that not even he would agree with explicitly. He argues for Poincaré's famous remark that "money is too important to be left to central bankers", concluding that the design of a rule could actually achieve a fair degree of monetary stability, that is, exactly what an independent central bank is designed to achieve, but is unable to (Friedman 2013).

## 3. The faintness of some arguments supporting the independence of central banks

### 3.1. The conservative central banker. A Prince Charming of the banking science

The contemporary rationale supporting central bank independence is the 'credible commitment' hypothesis. This theory posits that an independent central bank can more credibly commit to the long-term price stability goal, being less inclined to shock markets and agents by unexpectedly adjusting the money supply to leverage potential short-term advantages from these unforeseen changes. Barro and Gordon (1983) argued that a central bank endowed with the kind of credibility and reputation needed to influence the inflation expectations of economic agents could mitigate the characteristic inflationary bias of policymakers (see, e.g., Duffy and Heinemann (2021) doubts about the existence of a real-world central banker who can effectively manage the trade-off between credibility and flexibility). However, the omniscient, incorruptible governor, endowed with Robin Hood qualities in the fierce fight against inflation, is the creation of Rogoff (1985). Essentially, the Conservative banker is designed to substitute a rule. His aversion to inflation is enough for this. He is a law onto himself, the final authority, the head of an institution, which is, in itself, of last resort.

Here is a picture of the conservative central banker beyond economic good and evil: "A central banker cannot afford, in my opinion, to isolate in an ivory tower; on the contrary, it is necessary to pay constant attention to developments and trends in the economy and the financial system, as well as to the coordinates of government policies, in particular fiscal policy. This approach should not be assimilated to the promotion of discretionary conduct but is associated with a solid anchoring in everyday realities, whose complexity often exceeds the imagination of those who design economic rules or models" (Isărescu 2019, XXI). Here we have a sketch drawn by a central banker. Endowed with universal vision, imaginative enough to look beyond rules and models, capable, by himself, of solving any trilemma, endowed with monopoly. Fischer (1995) is seized by the idea of entrusting the monetary policy to a personality or to an institution. At the same time, we may be dealing with a dodge here. One also tried by Berger, de Haan, and Eijffinger (2001) in an attempt to align the Central Bank conservatism with that of an organization. However, behavioral attitudes target people, not the walls of an organization.

It is hard to believe that the modern literature on the topic of CBI relies on such arguments. It is hard to accept that it claims to be scientific by placing a human character as a source of objectivity in the reality of the financial world. In his famous treatise on economics, *Human Action*, Mises (1998) wanted to warn us, by the very title, that the objectivity of Economics lies in its subjectivity. Rogoff's impartial governor refuses this field of analysis. He has his own paradigm. One in which his aversion to inflation is put in econometric relation to his own pay (Persson and Tabellini 1993; Walsh 1993). We conclude and say that, under the shield of the Conservative governor, the idea of the CBI turns into a Prince Charming fantasy.

### 3.2. The Prophetic Principal-Agent Contract

It is convenient, it blends perfectly, for both the government and the Central Bank, that the relations between them be included in a contract like the one outlined by Walsh (1995). A contract grants rigor. Clearly defined rights, obligations and sanctions that govern the behavior of both parties. In our case, this is the idyllic part, the one that is intended to be shown.

Walsh's idea denounces the hypothesis that the state and the Central Bank, two public institutions, relate according to the rules of the game set by the private environment; that is, they are subject to the ruthless yoke of the free market, gratifying successes, sanctioning failures. Buchanan and Tollison (1972) showed that there is a political market in which governments wear out their opportunisms and selfish interests. In other words, there is no trace of the real market (Herrendorf 1998). History has proven that the bank was created to evacuate usurious behavior and sanitize the free market. The pro-independence supporters have proven something else. They have shown that the more independent the Central Bank is, the closer it is to the citizen; the more it is a state within a state, the better its chances of protecting the citizens from the endemic inflationary inclination of the usurper Principal.

Could we consider plausible the idea that two courts enjoying monopoly, one as a creditor and the other as a savior—as a last resort—engage in a game specific to private players? That they enter an Agency Contract, and the moral contingency and agency costs are the known ones? We think it is an illusion because:

First, no one has seen such a contract. We learn about its existence in the works of Walsh (1995), or Persson and Tabellini (1993). But the promoters of the idea of independence painted by the brush of the Principal-Agent contract do not reveal the contract. The Maastricht Treaty can only be considered at tentative level, with foggy initiatives in defining the Principal and the Agent. Otherwise, it is invisible because it is not convenient to be visible.

Accepted merely as an idea, the contract in question is doomed to be non-explicit, diffuse and directly non-binding. In this regard, Herrendorf (1998) identified a shortcoming in the incompleteness of Walsh's contract. The problem is that such a contract is a priori doomed to be incomplete, to stipulate only general clauses, "the exact details being left until a later date" (Coase 1937, 392). The main issue is that between these two pillars running the economy, the state and the Central Bank—although, for moments of public tenderness, they admit behaviors that are specific to the free market—they become "islands of conscious power in this ocean of unconscious co-operation" (Robertson 1923, 85) of the market, and an incomplete contract takes the form of an abnormal relation with the New Institutional Economics paradigm. Why is this so? Operating with the money of others, the Central Bank operates as an Agent to replace property with possession. It thus becomes not an absolute master but a very strong player as the holder of the right to control. By conceding indefinite rights, in residual form, the State is faced with an incomplete contract, which the Central Bank completes at will, identifying power with property. Thus, without even realizing it, it validates Grossman and Hart's idea (1986, 693): "we do not distinguish between ownership and control and virtually define ownership as the power to exercise control". The Central Bank does not distinguish between them either. And it does so until it gets to play both sides, being both Agent and Principal at the same time. The Central Bank emerged as state-owned banks with missions of general interest. These agents have never been elected; they were appointed. Establishing price stability as a starting point meant, from the beginning, an exploitable opportunity for the abuse of power (Jaillet 2019). Targeting prices was satisfactory just because it paved the way to the intimate structures of the economy; in fact, the goal proved unattainable. Thus, from the "children" of the government, the Central Banks became masters—not elected democratically—in areas where other political forces were democratically appointed. By handling "sub-objectives", the Central Bank came to deal with everything: money issuance, inflation targeting, security of the banking system, lender of last resort, economic growth, exchange rate, etc. Little of what is happening in the economy escapes the care and concern of the Central Bank. In all these areas, the Central Banks set their own operating rules. In other words, it is a Principal in disguise, yet a real one. The state is whipped. Its capacity as Principal is an illusion. Does this bother? Douglass North reassures us. Once in control of the money of others, like any other organization, the Central Bank can indulge in deforming its intentions because, he argues, "Institutions are not necessarily or even usually created to be socially efficient; rather they, or at least the formal rules, are created to serve the interests of those with the bargaining power to create new rules" (North 1990, 16). But when the "power to negotiate" has nothing to do with the market and is not subject to democratic exercise either, it is not difficult to guess where the CBI leads. It leads to a pale responsibility, difficult to enclose in a real contract. And when the contract is reduced to an unwritten agreement between the government and a governor, omniscient and a verified enemy of inflation, then the Principal-Agent contract becomes an insult; a paraphrase, in terms defined by the Central Bank to display something that does not exist.

Second, it is known that a real agency contract involves "agency costs" (Jensen and Meckling 1976). For the purposes of our analysis, we will only look into the "residual loss" that the Principal, the government in our case, should bear as a prosperity minus determined by an opportunistic behavior of the Agent. Does anyone account for such a thing? No! Because allegedly there is no residual loss on either the state, or on the population. That it does exist, however, is most easily seen in its maneuver to get rid of an awkward goal, to shift from price targeting to inflation targeting. It does know, a priori, that once prices rise, they do not fall back. On the contrary, they become "normal". The population pays, with a measurable loss of prosperity. Aiming at inflation, the Central Bank relegates this hypostasis—uncomfortable for its status—to history.

Overall, the alleged Principal-Agent contract is dilemmatic in practice but comfortable as an idea that floats in the air; not anyway, but endowed with the status of rule, of institution. Under its shield, the pay is double. First, by the fact that a contract grants creditworthiness. Second, a contract that is non-existent exempts from ungrateful responsibilities. Is this what Friedman must have had in mind when writing "I suspect that by far and away the two most important variables in their loss function are avoiding accountability on the one hand and achieving public prestige on the other" (as quoted in Fischer 1990, 1181)? In the battle for prestige, even an imaginary Principal-Agent contract turns out to be a good prop. Anyway, it is better than a rule. A rule obstructs the road to independence. When the area you are spreading your tentacles to reach the edge of the economy, a single rule does not suffice. In fact, the objective expression, the monetary-financial-fiscal mix, makes it difficult if not impossible to "sign" a Principal-Agent contract only between the government and the Central Bank. Many other Agents should participate. The Conservative governor would be losing its greatness.

### 3.3. Lender of last resort

The fact that such a function belongs, implicitly, to the structure of a Principal-Agent contract has become commonplace. Usually, the common phrase reads as follows: "The Central Bank assumes the responsibility of" this position.

The first step is to demonstrate that the Central Bank is at the core; that it is the supervisor, the monetary, if not the financial guardian. Prevention of endemic banking panics, deflation and recession is conceived in these terms (see, e.g., Freixas et al. 2000; Bernanke 2013) and this is also the source of a large share of its prestige.

Then, to become a lender of last resort, you must be, a priori, independent. A Principal-Agent contract run by the book does not speak of independence; it only defines the autonomy of the agent in managing the resources of the Principal, to its personal gain and with clearly described means. However, autonomy is not appropriate for the Central Bank's claims. It is only from the position of a highly independent institution that will you be able to oversee the entire banking system and produce macro effects through successive iterations of interest rates; and, as a last resort, save others as well; too big and too important to let them go bankrupt.

A third step consists in the formal omission of this function, as is the case with The Maastricht Treaty, which argues that the "natural" phase has already been reached. The lender of last resort function goes well with the issuing function. They belong to the intimate organics of the Central Bank. One must take them as criteria to define what is called a "fully-fledged" central bank. And a central bank becomes mature not upon some external request. No, "it is the way of thinking of the central bankers that determines whether a Central Bank has become, in fact, a lender of last resort" (Isărescu 2019, 82). A task taken on delightedly and emphatically because it grants prestige and, most definitely, it closes the proving circle of the need for independence. Does such a self-assignment raise any issues?

First, wielding the position of a lender of last resort is an attack on the spirit of the free market. The meaning of the free competitive market concerns all economic players. However, the rescue for insolvency problems or a temporary lack of liquidity only targets those on the nominal market. By discriminating, the bankruptcy of the institution of bankruptcy on the nominal side is encouraged. Second, there is the issue of the technology of the rescue process itself. Knowing *a priori* that you will be saved, even if this promise is not in writing, your behavior will not be in tune with the competitive spirit. Then, there is also the questionable issue of the source of the aid. If the Central Bank saves with the money of others and does not sanction irregularities, it is no longer a matter of moral contingency, but of serious immorality. More so as it allows itself to select the "chosen ones" only from the nominal area. Third, exerting this function perverts the content of the alleged Principal-Agent contract that it is supposed to be a part of. The lender of last resort belongs to the macro-prudential policy where prevention is more important than sanctions. The "Conservative governor" may not be fierce enough to deal with large-scale conflict situations where state presence is imperative. In this hybrid formula of "aid", the government may incur political costs, while the Central Bank gets all the glory (Müller 2019). The government comes out all wrinkled and the Central Bank carries the day. A perfectly twisted piece of evidence of an alleged Principal-Agent contract. Finally, what can one say about the brand-new role assigned to central banks, that of "dealer of last resort" (Mehrling 2010) for all assets on the financial markets, given that, they assured these were only temporary post-crisis measures, but which have turned out to be the new normal?

### 3.4. Financial autonomy on somebody else's money

If the instrumental, technical independence of the "dentist", as Keynes (1963, 373) called it, is undeniable and easy to capture in a causal relationship aimed at inflation, it is not as easy to find the link between the mentioned objective and the financial autonomy of the Central Bank.

Admittedly, financial independence is construed as a barrier to the potential predisposition to corruption of those who run a Central Bank. Financial independence appears here as a price for behavioral verticality. This idea, combined with the one derived from Article 107 of the Maastricht Treaty, leads to a strong logic: you can only afford to turn down advice or instructions provided you are financially independent. Here is the problem; it is not your money. Or, even if it is the bank of the banks, the Central Bank is and has the duty to remain the bank of the state. Lpo6Moreover, if that is the case, then the management of the funds it was endowed with must be subject to the same philosophy as in the case of public money.

According to an African tradition, anything beyond a seven-year lookback period becomes history. We are not interested in the beginnings; the Central Bank also argues. It is convenient to forget that at the beginning it received a dowry. Literature portrayed it as a "bride" endowed by the state. Any business starts either with its own funds or borrowed funds. This is not the case with a state bank. Its "groom" is generous. It would be interesting to find out what sources it used to pay its first salaries; how did it made the first expenses and if there was anything left for the development fund. Because the Central Bank itself states that it pays the following from its "own resources" obtained through profit accumulations. However, the facade overshadows the substance. When it declares that most of the seniority and profit goes to the state budget, it seems that the Central Bank is working for free, for patriotic reasons. It is not profit that defines its objectives. Its vocation is a civic, societal one. If one dares to analyze the exercise of obtaining and distributing the profit, one will learn something else: whatever happens, its financial independence is forever safe.

If those who have signed a payroll at a Central Bank remain, doctrinally, imbued with the sweetness of a generous pay and ready at any time to praise it, what do we make of the governor's pay? We have seen that to develop his divine inclinations and watch over his honesty, his salary is adjusted to the dynamics of inflation. What good are the models that prove the reverse proportionality between the two parameters when a recognized financially independent Central Bank sets its own salaries? Has anyone heard of the pecuniary punishment applied to the Boards of Directors or the governors of Central Bank during the 2008 crisis? We have not! What we do know is that the apparently "free lunch" forbidden to all is not forbidden to those empowered to manage it (Fischer 1995).

Ultimately, that is the problem. The alleged financial independence does not go well with equal responsibility. There are no sanctions. Here is what Fischer (2019, 331) disarmingly explains: "The most difficult issue is the sanctions that should be imposed on the Bank for failing to meet its targets … and there is no explicit sanction in most countries. Public reprimand and loss of reputation is probably a sufficient sanction". If the pain caused by the so-called humiliation had been so great, in 2008 we would have witnessed the mass resignations of Central Bank governors. What does it mean that this did not happen? First, a proof that it claims financial independence and reaching this goal has nothing to do with the salaries paid by banks. Was monetary policy completely innocent in the economic and financial turmoil of the mentioned crisis? Second, the Central Bank defies. No other state entity claims such a status. The Central Bank is allowed; accepting, most likely, a very "human" common sense: that you cannot touch mud without getting dirty! How could you deal with money and let others pay you?

### 3.5. The dilemmatic prohibition of budgetary financing

The ban on budget funding has been linked to the "temporal inconsistency" and the need to discipline governments. In other words, a democratically elected political power is relegated under the guidance of an *a priori* uncorrupted entity, specialized in the movement and rehabilitation of the world of money, to break the "spell of political influence" capable of inducing unfortunate situations.

What came out of this project? Conceived as a pro-independence mechanism, by weaning the state from its direct sources of funding, the result is perverted. The strong twinning between the monetary and financial renders the mechanism upside down, drives it to ask for cooperation and put the ban on budgetary funding between inverted commas. In addition, given the realities tangential to the project, it asks questions such as: Does the "temporal inconsistency" remain a convincing and sufficient argument to support the CBI and make it contrary to direct funding? Is the Central Bank kept within these frameworks, does it behave as required by this relationship? Does the government, in turn, live up to its commitment to self-discipline and refrain from brushing the independence of the Central Bank? Does the Central Bank remain an entity with a distinct game in the gear of an economy that demands systemic functionality?

First, it is worth noting how a significant number of reputed economists point out that the two major entities, the Central Bank and the government, trespass each other's territories, especially and conspicuously after the recent global crises that made the central bank far exceed its mandates (Coombs and Thiemann 2022); that the monetary-financial mix is obvious; that the extension of the Central Bank's job description is mostly fueled by the financial; that a dual responsibility of the Central Bank over the monetary and financial area should be subject to more thorough analysis; that the "temporal inconsistency" dilutes its explanatory force with the extension of the Central Bank towards the financial, etc. However, the positions are neutral; the Central Bank must remain independent. In this respect, Goodhart, Capie, and Schnadt (2019, 71), refer back to a 1959 episode to quote: "More than that, monetary policy, as we have conceived it, cannot be envisaged as a form of economic strategy which pursues its own independent objectives. It is a part of a country's economic policy as a whole and must be planned as such". Since 1959, the financial crises and the complexity of all complexities have asked for a different arithmetic for the monetary-financial relationship. Aglietta (1992) finds it appropriate to think of independence in terms of an "institutional arrangement", inviting the Central Bank to interact with the government and the markets. However, a perverse dynamic induced by finance can only be opposed to an independent bank, with a generous job description. From Trichet (2019), governor of the Bank of France, one learns that the function of bank of banks must be designed to ensure the stability of the monetary-financial system as a whole. As a case in point, in his country, France, it is not inflation or prices that are aimed at, but a "financial aggregate". However, warning that the Treasury could also be a last resort, the central banker sees independence as possible and necessary even in the conditions of a ubiquitous bank in the whole area of the money movement. Noticing the monetary-financial conjunction, Fischer (2019, 300), writes that "The most important source of institutional non-neutrality to inflation is the tax system; within the tax system it is the taxation of capital that is most distorted by inflation". The author thus notices that the two sectors interfere but does not seem impressed by a CBI that would extend to taxation. As early as 1994, Lamfalussy (2019) announced that financial innovation was invasive for monetary policy; that the current world phenomenon was financialization and not the monetization of the economy. As such, cooperation between the Central Bank and the Ministry of Finance is imperative. Volcker (2019) states that regardless of the degree of formal CBI, what matters is a mix of policies, "ideally a suitable co-ordination of policy". Who is in charge of such coordination? Not only over the monetary but also over the "fiscal policy, the labor market, social policies and other difficult issues"? The government? The Central Bank? The idea is that the monetary-financial system is an inseparable whole. In relation to this, what a Central Bank can aim at has nothing to do with any lesson of good practices to be delivered to the government. What it is left with is that within a jointly assumed "institutional arrangement", it has to interact with the government on the market. No chance of cutting it short when it is in need and asking for money.

Second, if "temporal inconsistency" is diluted as an argument for independence in the conditions of an increasingly obvious financial-monetary mix, other territories must also be explored. Or, other territories can be taken from the Ministry of Finance, which is not and cannot be as skilled as the Central Bank. Its unique access to credibility, the art of calibrating sophisticated models (why should they be sophisticated?), of elucidating far too complicated tasks, the wide, long horizon, and so on, these are all on its side—on the side of the Central Bank, not the Ministry of Finance. This is what Bernanke (2019) reassures us about. It remains to be seen how many long-term models coming out of the Central Bank's research labs are also sustainable.

Within the same philosophy, Debelle and Fischer (1994) argue that dynamic inconsistency is not the only reason for independence. In addition, the authors believe that the Central Bank should not be left to face political pressure alone. The "population's aversion" to inflation could be a way. Who should inspire them? By no means, the government. The Central Bank is much more qualified in this area as well. Equally interesting is what Professor Goodhart Charles says. A completely politicized bank does not seem possible. The government's desire not to completely get out of hand such a good milking cow ends up in a twisted outcome: "An 'independent' Central Bank will inevitably become much more politicized than a 'dependent' Central Bank" (Goodhart 1992, 33). Mugur Isărescu (2019, XL), the governor of the National Bank of Romania, also notices the wire dance of a bank that does not thrive in total independence, in a complete detachment "from the political consensus on appropriate actions". In the end, a protective umbrella could be necessary; a rescuing, or "mending" one. This makes it possible, in the fiery moments of the economy, for the bank to flee back home, leaving the government at gunpoint.

Recently, such "unconventional" situations have occurred; 2008, 2012 and post-Covid 19 are just three examples. Can one pretend not to know that budget deficits have become very "unconventional"? That Article 104 of the Maastricht Treaty went up in smoke, especially during the crises of 2008, 2012 and post-Covid 19? True, the Central Banks did not finance directly, but as banks of the banks, they did settle in the "logical" chain that leads indirectly to budgetary financing. The "cut-off" caused by the Covid 19 pandemic turned the "Stability Pact" into a great epsilon. Meanwhile, banks have been asked to raise money, with inflation control lagging. Moreover, it so happened that the Central Bank had to help their states get into debt, padding up their balance sheets with state securities. Tucker (2020, 1), a banker, admits that during the 2008, 2012 and post-Covid 19 crises, the Central Bank took back the "roles they used to have in the 1930s and 1980s, that is, simple tools in the service of the Ministries of Finance" (see also Mudge and Vauchez 2022). That is, they turned exactly into what the Treaty required them not to do.

In fact, the Central Banks detach from the institutionalized background and influence the debts of the states through something that is not visible at first glance. Through the way in which they handle interest rates, they influence the value ratios for foreign exchange. A high rate can make domestic goods more expensive than foreign ones. The result is that in an open economy, exports fall, and imports rise. The trade balance records the phenomenon and sends it to the balance of payments. It should also be noted that the dynamics of the interest rate is no stranger to the dynamics of the tax base (of turnover).

## 4. Loose environment, blurry outlines

A concept as assaulted as independence should come to the fore through defining benchmarks: a working definition, a commonly accepted taxonomy, clear objectives, intelligible mechanisms, etc. Unfortunately for the rigor required by acceptable knowledge, the ground on which it moves is slippery.

The exercise begins by not identifying a working *definition* of independence. One discovers that independence cannot exist in its pure state; that it can be political, economic, real, or legal; institutional-strategic or tactical; formal or informal; within or against the state; neither too big nor too small, and so on, and so forth. How many of these invoked forms of manifestation are directly related to the underlying motivation, that of "temporal inconsistency"? Fancy being a Central Bank and be allowed to arrogate any of the above adjectives.

The exercise of the most "scientific" gobbledygook in Economics continues with the search of the *goal*. Theoretically, the Central Bank is granted independence to deal freely with ensuring and maintaining price stability. From theory to practice, the road is sprinkled with sub-objectives, targets, goals, functions, and purposes which, most of the time, the Central Bank takes on. However, the place where the CBI seems to be looking for credible contours, but fails to convince, is that of the shifting sands on which *targets* are set. Prices, inflation, money supply or all in one place? Due to the "temporal inconsistencies", the Central Bank is delegated to price stability. But the Central Bank knows how to give a favorable interpretation to the context to facilitate its mission and dissipate its responsibility. By origin, there should be no other goal than price stability. In fact, even though this does not seem to be a breach of the Principal-Agent contract, inflation targeting seems more appealing. At the level of common understanding, the difference in adjectives is related to ornaments. Does the Parliament or the citizens know why the switch is changed? As we have already noted, price targeting involves the coordination of economic processes whose articulation and chaining cannot be the sole responsibility of the Central Bank. All economic players related to the supply-demand mechanism participate in this process. To bring an already high price level back to the starting point, one needs to be more than just a master in handling the interest rate. Other influences and other determinants may be stronger in worsening the supply-demand tensioned relation, so that the end flows towards restoring prices to the starting point. The structure and size of supply or demand, civil or non-civilian in nature, the intensity of agency in a global economy, investment promotion and commissioning policy, the length of production periods, the state of trade balance or policy payroll, are just some of the aspects related to price dynamics, regulated by the governments, not by the Central Banks. The chance of damaging its status by taking on such a complicated task, although very necessary for the health of an economy, is very high. It is less risky and more convenient to target inflation. If inflation goes beyond the target and prices rise accordingly, the Central Bank will only adjust its instruments. What's done is done, could happen again; prices remain where they were driven by exceeding the target. There is no need for sophisticated calculations to restart the process. If calculations are conducted, this is to induce the idea that the interest rate is not fixed, it results from calculations. After all, the very conservative governor wears the coat of the well-known *comissaire-priseur* from the Walrasian general equilibrium model, which, through successive iterations, testing the market, reaches an equilibrium price. The nuance is that the governor is interested in the equilibrium price of money. That is what we think Goodhart, Capie, and Schnadt (2019, 51) mean when they write that "central bankers need to vary interest rates in response to deviations of the uncertain future rate of inflation (from the desired, say, 0 to 2 per cent rate), rather than react to current data". The "current data" consists of prices, production, and employment. They stay where they were!

What takes the issue into derision is the way in which this dilemma, or trilemma, is bathed in all waters: prices-inflation-money supply. Goodhart, Capie, and Schnadt (2019, 99) believe that price stability "leaves open a great deal of flexibility about the choice of immediate, specific, short-term objectives". Under the guise of price stability are listed all the sub-objectives that the Central Bank chooses at will. The chairman of the Fed, Alan Greenspan (2019, 263) believes in the "guiding principle" of the Central Bank: "Stability of all prices, including assets and financial stability". The French banker Trichet (2019, 269) does not seem to think so: "prices, interest rates and exchange rates cannot take the place of a money supply growth target" and adds: "Tracking the ultimate goal of price stability alone raises the danger that the central bank will not be able to anticipate trouble". The unshakable "belief in price stability" remains the leitmotif of Japanese banker Yasushi Mieno (2019, 272). Moreover, Fischer (2019, 311) ends up saying that "My present view is that the inflation target with its greater short-term inflation rate certainty is preferable, despite its greater long-term price level uncertainty". Of course, especially in the long-term, price targeting can bring clouds to the clear sky of the Central Bank. For the peace of the Central Bank, Alexandre Lamfalussy (2019, 365) finds another way: "Money supply targeting relieves central banks of some of the pressure which might be exerted on them by governments or parliaments". Anything but to answer to the "king" about how the central bank serves the "public welfare". The German banker Karl Otto Pöhl (2019, 381) states that if we only discuss the stability of prices, and we do not tackle the stability of the financial system "whether we can declare victory, remains to be seen" (see also Svensson 1996; Guender and Do Yoon Oh 2006). Fullwiler and Wray (2013) argue that "beguiled" by the low levels of inflation over the past two decades, the Fed believed its policies were working well, until the fatal outcome: the financial crisis of 2008.

Whom to believe? The problem is that in such a field, foggy and devoid of clearly emphasized rules and concepts, one does whatever wants of independence. In fact, as Pöhl (2019) argues, price stability alone does not bring victory; the territory must be expanded. No effort is spared to find out what is left of the Philips curve, via Samuelson and Solow, Friedman, Phelps and Lucas. In vain did Lucas solve the problem. Other exploitable surroundings must be revived or identified. The "average" rigidity of wage contracts—the dispute Fischer (1977) *versus* Sargent and Wallace (1975)—causes some trouble until one learns that these contracts are not signed at once, nor are they terminated at once; that monetary expansion on this way is problematic, the tidal effect does not occur. Yet, proof must be provided that every fraction of the interest rate is worked out, that nothing is random, it is all thorough calculation. The solution resides in the DSGE models (see, e.g., the COMPASS model of the Bank of England—Burgess et al. 2013; the FRB/US model of the Fed—Brayton, Laubach, and Reifschneider 2014).

Moreover, institutional laxity, the lack of rules or the existence of questionable rules can only support, in turn, a questionable independence. What to choose from the inventory of rules kept in the drawer? Friedman's rule of the dynamics of the money supply in relation to the dynamics of GDP? The long-term currency neutrality rule? Walsh's rule of a Principal-Agent contract? Rogoff's rule—a conservative central banker is the key to fighting inflation? Keynes' rule—the CBI is that of the dentist? The Maastricht rule—we do not receive instructions and advice from anyone?

That being the case, the chance of the discretionary to prevail over the rule is assured. In this regard, apprehensive to the fast pace of change in free market structures, Goodhart, Capie, and Schnadt (2019, 103) conclude that "central banks will rightly aim to retain their discretionary flexibility", to adapt to circumstances. Unrestricted freedom to "circumstances"! In the same vein, Greenspan (2019, 266-267) argues that a "predetermined path", a rule, is an illusion. He believes that "there are no simple models that can be inferred out of some complex set of econometric techniques that can provide a definitive guide for monetary policy, or for that matter economic policy in general". Conclusion: "There is clearly no rigid distinction between model-based 'discretionary policy' and those policies that are on automatic pilot with a specific target in mind. The latter is a version of the former". Here it is revealed the futility of scholarly econometric exercises summoned, apparently, to objectify the decisions of the Central Bank. What results thereof, states someone from the heart of the system, resembles discretionary policy. As discretionary as the one that the conservative governor of a bank sets "in mind". Does it take independence to reach such a conclusion?

## 5. Conclusion

To defend the territory naturally recognized as one's own, there is no need for theories sent beyond the natural edges of the perimeter. However, a theoretical outpouring of such proportions says something else.

Firstly, the Central Bank's great battle is against the Ministry of Finance. In fact, it is not the break from the Ministry of Finance that concerns it but subordinating and disciplining it. Or, these entities' attempt to discipline one another, since at least at instrumental level they are deemed to cooperate, is an improbable idea and an impossible practice. The impression that one entity disciplines the other through interest rates and non-financing, while the other, compliantly minds its own budgetary optimum, with no connection to the monetary optimum, is an illusion.

Secondly, the rigor of the econometric exercise it resorts to provide rigor to the analysis, gives it the upper hand, but that is all. As Alan Greenspan argues, the whole laborious mathematical process does not rule out the possibility that the interest rate be fixed in the governor's office. In addition, if in the long run, independence is not related to the real dynamics of the economy, as the validation attempts show, the mathematical exercise occurs in a vacuum.

Thirdly, another noteworthy aspect of the problem derives from the fact that, very often, the CBI stems from an unfriendly relationship with democracy. In such a country, only the citizen is free. Everything is geared toward the citizen, including the energies burnt in Parliament, the Government, or the Presidency and, why not, in the Central Bank. The optimal and monetary budget is designed for the citizen. In vain is the country's treasury full if the citizen is not well.

Fourthly, the issue of independence is also a matter of ownership. The Central Bank's money is actually someone else's money; to an overwhelming extent, it is the state's money. The imaginary Principal-Agent contract does not imply a transfer of ownership. Thus, if it manages a collective pocket, the Central Bank has to give account, to the point, professionally, to the Parliament. Liability involves the existence of brake pads, rules. Or, in this case, the Central Bank proves very tenacious in convincing that circumstantial, institutional laxity, define normality. The exception resides in the infatuated rule: the permission to refuse any advice or instruction from the political power.

Finally, the rejection of the free-market philosophy is one of the great insults that the CBI makes to an open society. No one denies the normative character of the Central Bank; only not the way it perceives itself. The economy, in which it is invited to manifest, is more natural and thus rejects its epic tales of the deified governor, as well as the speculations fixed on the imaginary of pure economies. By denying the free market its episodes of spontaneity and directing the way it should behave, saving its failed products to make good use of their money, the Central Bank, despite its independence, places itself in the imaginary, and becomes independent and master of an improbable world.